\documentclass[runningheads]{llncs}
\usepackage[T1]{fontenc}
\usepackage{algorithmic}
\usepackage{comment}
\usepackage{graphicx}
\usepackage{textcomp}
\usepackage{xcolor}
\usepackage{multirow}
\usepackage{booktabs}
\usepackage{wrapfig}
\usepackage{caption}
\usepackage{colortbl}
\usepackage{subcaption}

\usepackage{amssymb}
\usepackage{amsmath}
\usepackage{hyperref}
\usepackage{xspace}
\usepackage{subcaption}

\newcommand{\etal}{\textit{et al}\xspace}

\usepackage[T1]{fontenc}
\usepackage{graphicx,verbatim}
\begin{document}
\title{UD-Mamba: A pixel-level uncertainty-driven Mamba model for medical image segmentation}

\author{Weiren Zhao \textsuperscript{1} \and
Feng Wang \textsuperscript{2} \and
Yanran Wang \textsuperscript{3} \and
Yutong Xie \textsuperscript{4} \and \\
Qi Wu \textsuperscript{5} \and
Yuyin Zhou \textsuperscript{1}}

\authorrunning{W. Zhao et al.} % 如果作者超过两个，使用第一作者et al.的形式

\institute{
\textsuperscript{1} University of California, Santa Cruz, CA, USA\\
\textsuperscript{2} Johns Hopkins University, Baltimore, MD, USA \\
\textsuperscript{3} Stanford University, Stanford, CA, USA \\
\textsuperscript{4} Mohamed bin Zayed University of Artificial Intelligence, Abu Dhabi, UAE \\
\textsuperscript{5} University of Adelaide, Adelaide, SA, Australia 
}

\maketitle

\begin{abstract}

Recent advancements have highlighted the Mamba framework, a state-space model known for its efficiency in capturing long-range dependencies with linear computational complexity. While Mamba has shown competitive performance in medical image segmentation, it encounters difficulties in modeling local features due to the sporadic nature of traditional location-based scanning methods and the complex, ambiguous boundaries often present in medical images. To overcome these challenges, we propose \textbf{U}ncertainty-\textbf{D}riven \textbf{Mamba (UD-Mamba)}, which redefines the pixel-order scanning process by incorporating channel uncertainty into the scanning mechanism. UD-Mamba introduces two key scanning techniques: 1) \textbf{sequential scanning}, which prioritizes regions with high uncertainty by scanning in a row-by-row fashion, and 2) \textbf{skip scanning}, which processes columns vertically, moving from high-to-low or low-to-high uncertainty at fixed intervals. Sequential scanning efficiently clusters high-uncertainty regions, such as boundaries and foreground objects, to improve segmentation precision, while skip scanning enhances the interaction between background and foreground regions, allowing for timely integration of background information to support more accurate foreground inference. Recognizing the advantages of scanning from certain to uncertain areas, we introduce four learnable parameters to balance the importance of features extracted from different scanning methods. Additionally, a cosine consistency loss is employed to mitigate the drawbacks of transitioning between uncertain and certain regions during the scanning process. Our method demonstrates robust segmentation performance, validated across three distinct medical imaging datasets involving pathology, dermatological lesions, and cardiac tasks. Code is available at \href{https://github.com/piooip/UD-Mamba}{https://github.com/piooip/UD-Mamba}.


\end{abstract}
\section{Introduction}
Transformers have shown significant potential in image processing due to their ability to model long-range dependencies~\cite{NIPS2017_3f5ee243,dosovitskiy2020image,liu2021swin,bao2024channel,zhang2024synergistic}. However, their quadratic computational complexity with respect to sequence length imposes substantial computational costs, particularly in high-resolution tasks like medical image segmentation. Recently, state-space models (SSMs) have emerged as a more computationally efficient alternative, offering linear complexity while preserving the ability to model long-range dependencies~\cite{gu2021efficiently}. Among these, the Mamba architecture~\cite{gu2023mamba,dao2024transformers} stands out, employing selective scanning techniques and hardware-optimized design to achieve impressive results across various visual tasks~\cite{liu2024vmamba,zhu2024vision,li2024videomamba,hu2024zigma,liu2024swin}.

\begin{figure}
    \centering
    \includegraphics[width=1\linewidth]{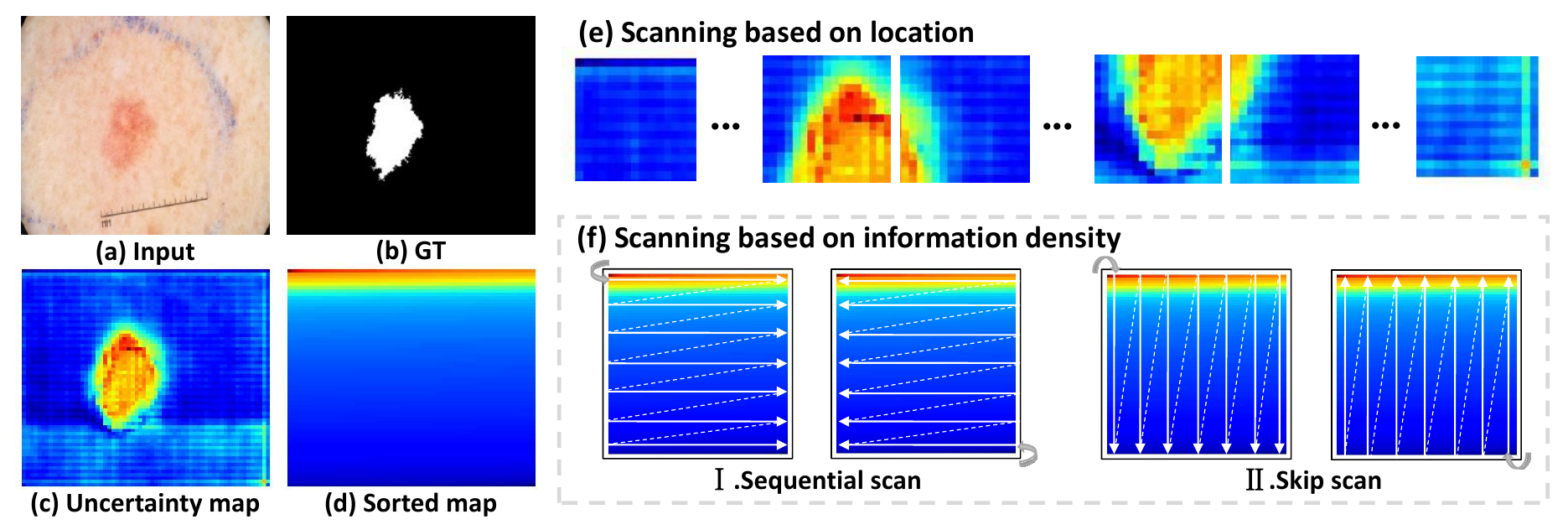}
    \caption{Pixel-level channel uncertainty-based scanning mechanism. (a) Input image; (b) Ground truth; (c) Resulting image obtained from channel-based uncertainty calculations, where pixels with the highest uncertainty are red and the lowest are blue; (d) Feature image sorted by the degree of uncertainty; (e) Previous method using the SS2D~\cite{liu2024vmamba} scanning mechanism; (f) Our UD-SSM scanning mechanism, which includes sequential scanning and skip scanning.}
    \label{fig:scan}
\end{figure}

In medical image segmentation, the primary objective is to accurately delineate regions that correspond to target organs or pathological tissues, providing essential support for clinical diagnoses~\cite{ronneberger2015u,chen2024transunet,isensee2021nnu,hatamizadeh2022unetr,li2018h,wang2021automatic}. Due to its capacity to capture long-range dependencies and process high-resolution images efficiently, the Mamba framework has seen increasing application in the medical imaging field~\cite{yang2024mambamil,xing2024segmamba}. However, Mamba’s traditional position-based sequential scanning method often leads to intermittent scanning of different semantic regions (\autoref{fig:scan}(e)), which is particularly problematic when dealing with complex backgrounds and ambiguous boundaries in medical images. This hinders Mamba’s ability to accurately model local features essential for effective segmentation~\cite{fan2024slicemamba,wang2024mamba}.

Based on the effectiveness of previous uncertainty methods in medical image segmentation~\cite{xia2020uncertainty,shi2021inconsistency,mehrtash2020confidence,jungo2019assessing,baumgartner2019phiseg}, we propose Uncertainty-Driven Mamba (UD-Mamba) to overcome this limitation. The core of UD-Mamba is the Uncertainty-Driven Selective Scanning Model (UD-SSM), which leverages channel uncertainty as a guiding metric to redefine the pixel-wise scanning process. As illustrated in~\autoref{fig:scan}(c), pixels with higher median channel uncertainty are often associated with critical areas, such as the foreground and boundaries. Conversely, regions with lower uncertainty are typically related to the background. By calculating the uncertainty map and ranking the pixels based on their uncertainty levels, as shown in~\autoref{fig:scan}(d), we ensure that uncertain (and thus critical) regions are distinguished from more certain regions (typically representing background information).

\begin{figure}[htbp]
    \centering
    \begin{minipage}[b]{0.45\linewidth}
        \centering
        \includegraphics[width=\linewidth]{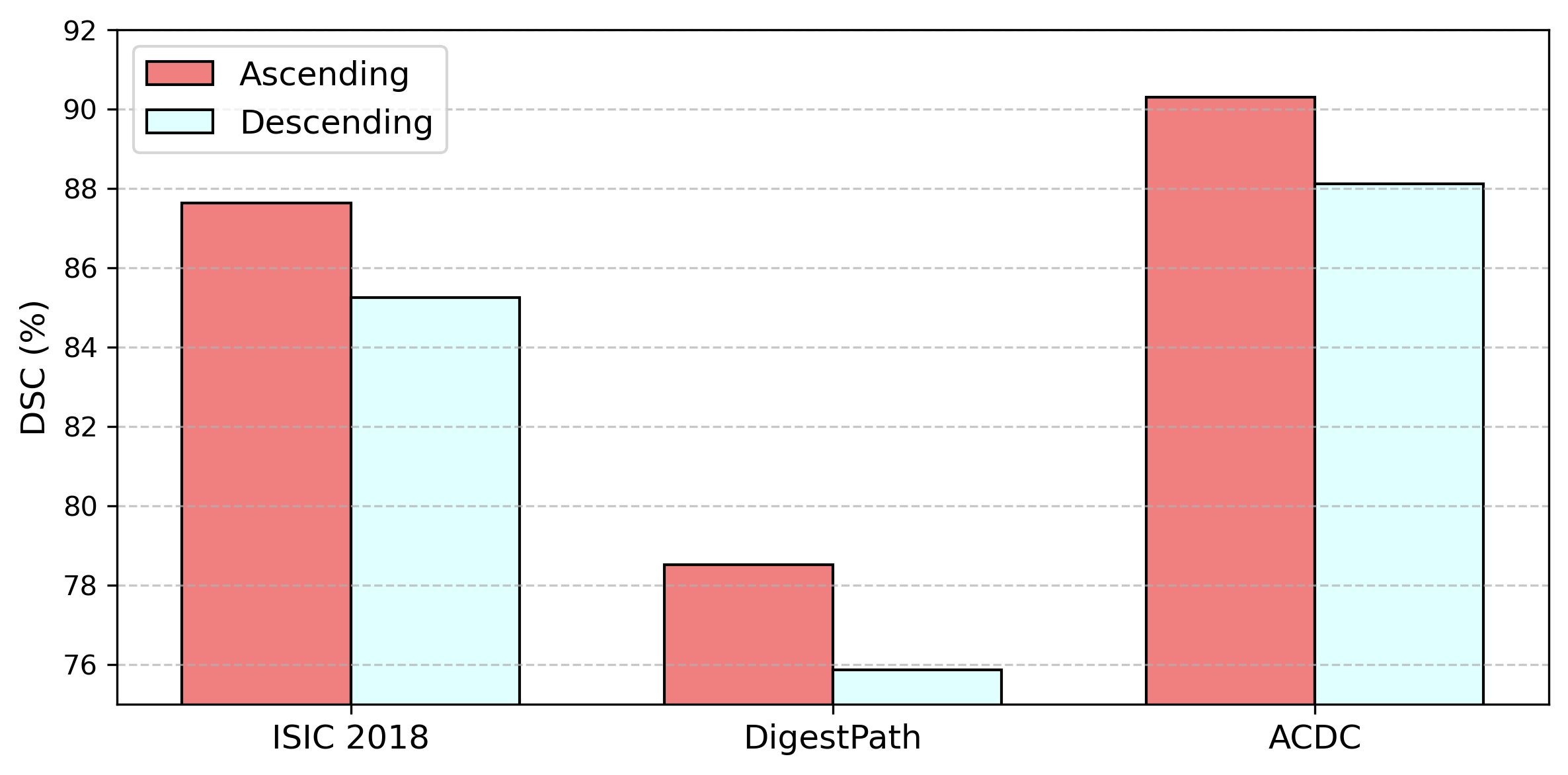}
        \caption{Performance comparison of ascending and descending methods: the ascending method scans from regions of low uncertainty to high uncertainty, while the descending method scans in the reverse order.}
        \label{fig:compare}
    \end{minipage}
    \hfill
    \begin{minipage}[b]{0.5\linewidth}
        \centering
        \includegraphics[width=\linewidth]{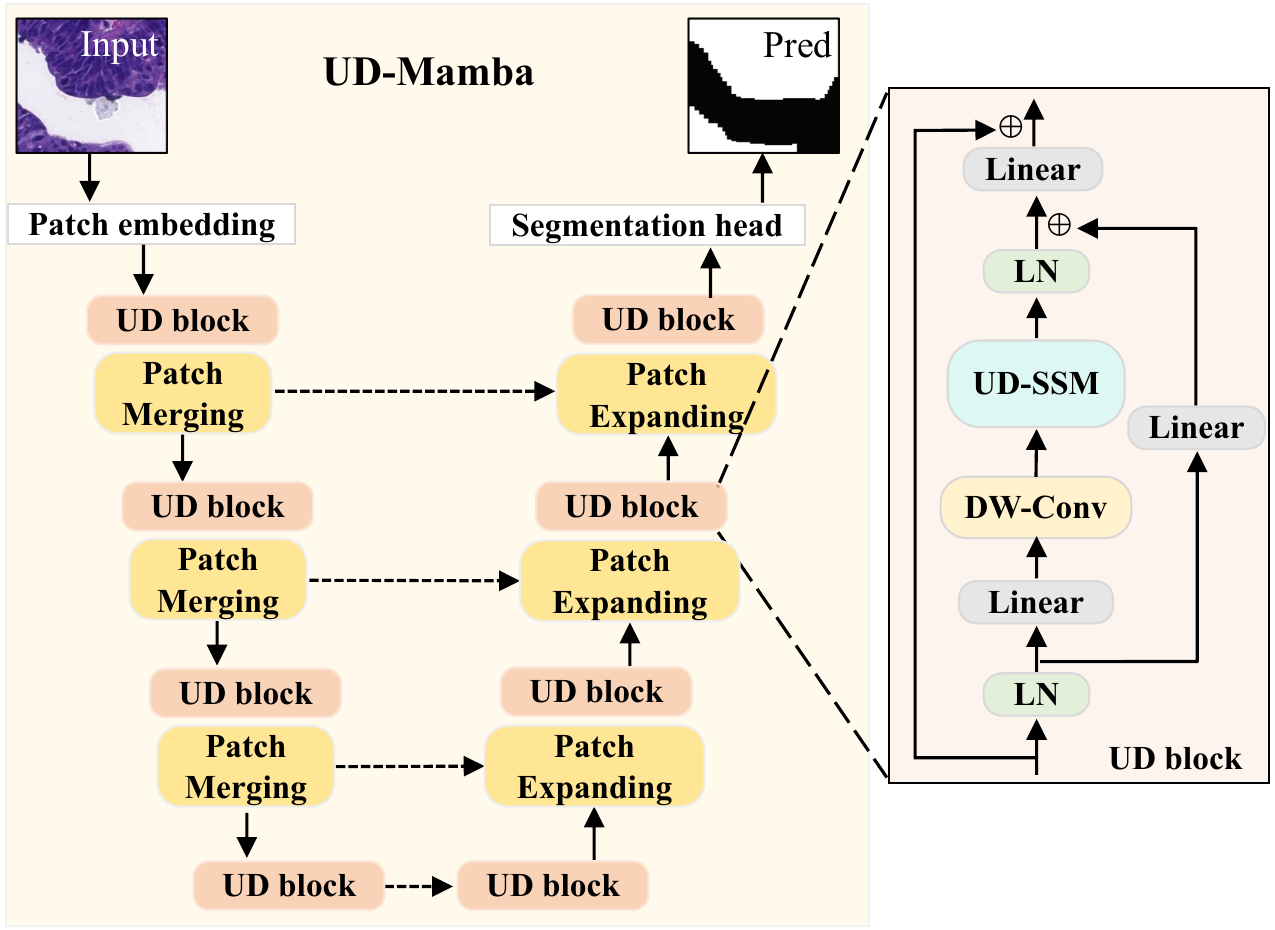}
        \caption{Illustration of the UD-Mamba architecture, which includes a patch embedding layer, an encoder-decoder with Uncertainty-Driven (UD) Blocks, and a segmentation head. Each UD Block features the Uncertainty-Driven Selective Scanning Model (UD-SSM) for processing input.}
        %Illustration of our UD-Mamba architecture, featuring a patch embedding layer, an encoder-decoder with Uncertainty-Driven (UD) Blocks, and a segmentation head. Each UD Block incorporates the proposed Uncertainty-Driven Selective Scanning Model (UD-SSM) as a core component and processes input through a series of operations.}
        \label{fig:UD-mamba}
    \end{minipage}
    \label{fig:comparison}
\end{figure}

The proposed scanning strategy in UD-SSM, as depicted in~\autoref{fig:scan}(f), includes two key methods: 1) \textbf{Sequential scanning}: This method processes pixels in strict order according to their uncertainty levels, effectively clustering high-uncertainty regions such as boundaries and foreground areas. By focusing on these critical regions, sequential scanning ensures that the model captures the fine details in areas crucial for accurate segmentation. 2) \textbf{Skip scanning}: This technique moves vertically across the image at consistent uncertainty intervals, enhancing the interaction between background and foreground information. It supplements the model's understanding of background regions while ensuring precise foreground segmentation. By combining sequential and skip scanning, UD-Mamba is able to focus on the fine structures of critical regions while maintaining an understanding of the broader context. This dual-scanning approach enables a more balanced and effective segmentation performance. 
Inspired by Vision Mamba~\cite{zhu2024vision}, we further introduced forward and backward scanning mechanisms in Sequential scanning and Skip scanning. However, we observed that scanning from low uncertainty regions to high uncertainty regions generally yields better results than reverse scanning, as shown in \autoref{fig:compare}. To optimize this process, we introduce four learnable parameters that adjust the importance of features gathered from different scanning techniques. Additionally, we apply a cosine consistency loss to ensure that features derived from scanning high-to-low uncertain regions are aligned with those from low-to-high uncertain regions, further enhancing segmentation accuracy.

%\begin{figure}
%    \centering
%    \includegraphics[width=1\linewidth]{high-low3.png}
%    \caption{Performance comparison of ascending and descending methods: the ascending method scans from regions of low uncertainty to high uncertainty, while the descending method scans in the reverse order.}
%    \label{fig:compare}
%\end{figure}

%\begin{figure}
%    \centering
%    \includegraphics[width=0.6\linewidth]{UD-Mamba1.pdf}
%    \caption{Illustration of our UD-Mamba architecture, featuring a patch embedding layer, an encoder-decoder with Uncertainty-Driven (UD) Blocks, and a segmentation head. Each UD Block incorporates the proposed Uncertainty-Driven Selective Scanning Model (UD-SSM) as a core component and processes input through a series of operations.}
%    \label{fig:UD-mamba}
%\end{figure}

Our contributions can be summarized as follows:
\begin{itemize}
    \item We introduce a novel pixel-level selective scanning approach guided by channel uncertainty, addressing the limitations of traditional position-based sequential scanning methods.

    \item We incorporate learnable parameters to balance feature importance across different scanning directions and employ a cosine consistency loss to align forward and backward scan results, improving feature consistency.

    \item Extensive experiments on three medical imaging datasets demonstrate that UD-Mamba effectively identifies ambiguous regions, leading to more reliable segmentation outcomes compared to existing Mamba-based methods.
\end{itemize}

\label{sec:intro}
\section{Related Work}

\subsection{Medical Image Segmentation}
Medical image segmentation refers to the process of segmenting medical images into dense predictions of pixels corresponding to lesions or organs based on imaging methods such as CT ~\cite{zhou2019semi,zhou2019prior,wang2019abdominal,fu2021review} and MRI~\cite{ji2022amos,zeng2020review}. Among them, Convolutional Neural Networks (CNNs) and Transformers dominate as leading frameworks. A significant advancement in CNN-based segmentation was introduced by UNet~\cite{ronneberger2015u}, which employs a symmetric encoder-decoder architecture with skip connections. These skip connections effectively integrate local features from the encoder with semantic information from the decoder, setting the foundation for many subsequent improvements~\cite{zhou2019unet++,oktay2018attention,le2023rrc,huang2021missformer,tang2021recurrent}. Despite its success, CNN-based methods are limited by their local receptive fields, which hinder the capture of long-range dependencies essential for dense prediction tasks.

Inspired by the Vision Transformers (ViTs)~\cite{dosovitskiy2020image}, there has been increasing interest in incorporating Transformers into medical image segmentation~\cite{hatamizadeh2022unetr,zhou2023nnformer,lin2022ds,huang2022missformer,wang2022mixed,zhao2024semi}. TransUNet~\cite{chen2024transunet}, one of the pioneering works, introduced a hybrid model that uses Transformers in the encoder to model global context, while retaining the overall UNet structure. Swin-UNet~\cite{cao2022swin} further explored a fully Transformer-based framework for segmentation tasks. While Transformers are adept at modeling long-range dependencies, their self-attention mechanism introduces quadratic complexity relative to input size, which poses scalability challenges, especially in pixel-level tasks like medical image segmentation.

\subsection{State space models for segmentation}
State Space Models (SSMs) have recently emerged as a powerful tool for visual tasks, with Mamba~\cite{gu2023mamba,dao2024transformers} showing promising results by efficiently modeling global context with linear complexity. Mamba-based models have demonstrated their versatility across a range of applications~\cite{zhu2024vision,ruan2024vm,he2024mambaad,zhang2024voxel,fan2024slicemamba}. U-Mamba~\cite{ma2024u} introduces a hybrid framework combining CNNs and SSMs, effectively capturing both local and global features. Swin-UMamba~\cite{liu2024swin} incorporates ImageNet-based pretraining into a Mamba-based UNet for enhanced medical image segmentation performance.
P-Mamba~\cite{ye2024p} combines Perona-Malik diffusion with Mamba to improve echocardiographic left ventricular segmentation in pediatric cardiology. Additionally, Wang \etal.~\cite{wang2024large} introduced LMa-UNet, a Mamba-based network with a large-window design for improved global context modeling.

Despite these advances, accurately segmenting complex medical images remains a challenge due to the intricate background and ambiguous class boundaries. Moreover, traditional scanning mechanisms, which intermittently scan different semantic regions, limit the model's ability to consistently capture the full range of contextual information within the images.

\subsection{Uncertainty estimation in segmentation}
In recent advances in uncertainty estimation for medical image segmentation, various methods have highlighted the importance of incorporating uncertainty to enhance model reliability and performance~\cite{lu2023uncertainty,wei2023consistency,monteiro2020stochastic,wang2019aleatoric,zheng2021rectifying,fan2022ucc}. Wang \etal.~\cite{9710267} proposed a domain-adaptive segmentation framework that refines pseudo-labels with uncertainty awareness, reducing the impact of noisy labels. Similarly, Zhang \etal.~\cite{zhang2023uncertainty} introduced an uncertainty-guided mutual consistency learning framework, leveraging estimated uncertainty to select reliable predictions in semi-supervised segmentation. Li \etal.~\cite{li2023region} employed evidence-based deep learning (EDL), focusing on a region-based EDL framework that utilizes Dempster-Shafer theory to deliver robust segmentation outcomes with quantifiable uncertainty. Collectively, these approaches underscore the role of uncertainty estimation as a critical factor in enhancing the trustworthiness and clinical applicability of medical image segmentation models.

\label{sec:related}

\section{Method}

In this section, we begin with a description of the preliminaries of selective SSM~\cite{gu2023mamba}, a foundational concept central to the Mamba framework. Next, we provide a comprehensive overview of our proposed UD-Mamba architecture, with an overall framework illustrated in \autoref{fig:UD-mamba}. %We highlight that the proposed UD-Mamba framework incorporates the powerful UNet architecture~\cite{ronneberger2015u}. 
Finally, we elucidate the key components of UD-Mamba, detailing the operational workflow of the Uncertainty-Driven Selective Scanning Model (UD-SSM) shown in \autoref{fig:CD-Block} and the derived optimization strategies.

\subsection{Preliminaries}
% In Mamba blocks, the token mixer operates as a specialized selective SSM, which is characterized by its efficient handling of long-range dependencies through a compact memory representation. The model defines four core input parameters \((\Delta, \mathbf{A}, \mathbf{B}, \mathbf{C})\), which are transformed into \((\overline{\mathbf{A}}, \overline{\mathbf{B}}, \mathbf{C})\) using the following state-space dynamics:
% \begin{equation}
% \begin{gathered}
% \overline{\mathbf{A}}=\exp (\Delta A) \\
% \overline{\mathbf{B}}=(\Delta \mathbf{A})^{-1}(\exp (\Delta \mathbf{A})-\mathbf{I}) \cdot \Delta \mathbf{B}
% \end{gathered}
% \end{equation}
% The Mamba block excels in efficiently modeling temporal sequences using a structured state-space representation. The sequence transformation in the SSM is expressed as:
% \begin{equation} \label{eq:y_t}
% \begin{gathered}
% h_t = \overline{\mathbf{A}} h_{t-1} + \overline{\mathbf{B}} x_t \\
% y_t = \mathbf{C} h_t
% \end{gathered}
% \end{equation}
% Here, \(t\) refers to the temporal index, \(x_t\) is the input sequence at time \(t\), \(h_t\) is the hidden state capturing the temporal context, and \(y_t\) represents the output. The hidden state \(h_t\) serves as a compact, memory-efficient repository that retains essential historical information, allowing the model to propagate context across time steps without increasing computational burden.

The Mamba block leverages a specialized Selective Scan Mechanism (SSM) designed to efficiently handle long-range dependencies by maintaining a compact memory representation. Inspired by the Kalman filter, this SSM functions as a linear time-invariant (LTI) system~\cite{kalman1960new}, transforming input parameters \((\Delta, \mathbf{A}, \mathbf{B}, \mathbf{C})\) into \((\overline{\mathbf{A}}, \overline{\mathbf{B}}, \mathbf{C})\) through a structured state-space formulation. The continuous-time dynamics are given by
\begin{equation}
\begin{gathered}
\overline{\mathbf{A}} = \exp (\Delta \mathbf{A}) \\
\overline{\mathbf{B}} = (\Delta \mathbf{A})^{-1}(\exp (\Delta \mathbf{A}) - \mathbf{I}) \cdot \Delta \mathbf{B}
\end{gathered}
\end{equation}
In tackling the limitations of traditional LTI SSMs in capturing contextual information, the Mamba block employs an input-dependent selection mechanism (referred to as S6) that adapts dynamically to input variations. Its recursive relation is expressed as
\begin{equation} \label{eq:y_t}
\begin{gathered}
h_t = \overline{\mathbf{A}} h_{t-1} + \overline{\mathbf{B}} x_t \\
y_t = \mathbf{C} h_t
\end{gathered}
\end{equation}
where \(t\) is the temporal index, \(x_t\) is the input sequence at time \(t\), \(h_t\) represents the hidden state capturing temporal context, and \(y_t\) is the output. By employing associative scan algorithms with linear complexity, the Mamba block efficiently computes responses, allowing context propagation across time steps while minimizing computational burden.

\subsection{UD-Mamba}\label{sec:3.2}

The UD-Mamba architecture leverages a streamlined yet robust UNet framework~\cite{ronneberger2015u}, as illustrated in~\autoref{fig:UD-mamba}. It comprises three key components: a patch embedding layer that transforms the input image into a sequence of patches for subsequent processing, an encoder-decoder structure composed of Uncertainty-Driven (UD) Blocks that captures and integrates both local and global features across varying scales, and a segmentation head that produces the final pixel-wise segmentation output based on the decoded features. 
The encoder-decoder configuration is enhanced by skip connections, which facilitate the integration of multi-scale feature representations. This architectural choice enhances information propagation across levels, ultimately improving segmentation accuracy.

Each UD Block processes input through a sequence of operations. The input is first transformed via layer normalization (LN) and passed through a linear layer. Next, depthwise convolutions (DW-Conv) followed by a SiLU activation function are applied. The data is then processed by the UD-SSM module, which captures long-range dependencies with linear complexity using proposed sequential and skip scanning strategies. A residual connection combines the UD-SSM output with earlier features, followed by additional refinement through a final linear layer.

\begin{figure*}
    \centering
    \includegraphics[width=1\linewidth]{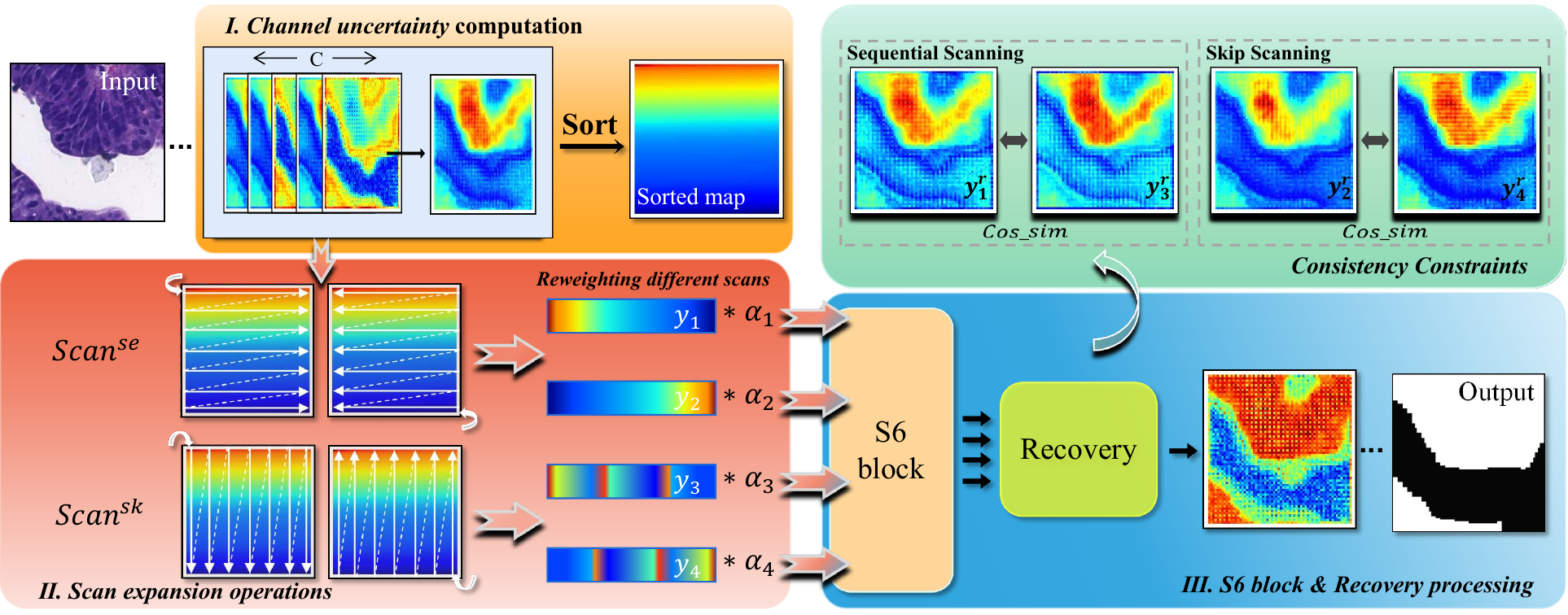}
    \caption{Detailed description of the UD-SSM. I. Describes the uncertainty calculation process based on channel uncertainty in the UD-SSM.  II. Explains the scanning expansion operation, which mainly includes two strategies: Sequential scanning (\(\text{Scan}^{se}\)) and Skip scanning (\(\text{Scan}^{sk}\)), as well as the subsequent Reweighting operation. III. Presents the S6 block and Recovery processing, while also introduces the calculation process of Consistency Constraints.}
    \label{fig:CD-Block}
\end{figure*}

\subsection{UD-SSM}
Traditional SSMs struggle with effectively modeling local features due to intermittent scanning of target regions, we propose a pixel-level uncertainty-driven selective scanning approach, UD-SSM. This method is distinct from conventional pixel-order scanning mechanisms, as it leverages uncertainty at the pixel level to inform scanning sequences. As illustrated in~\autoref{fig:CD-Block}, our UD-SSM introduces the following key components: channel uncertainty computation, scan expansion operations, the S6 block~\cite{gu2023mamba} and the recovery processing.

\noindent \textbf{Channel uncertainty computation}.
Given an input feature tensor \( \mathbf{X} \in \mathbb{R}^{C \times H \times W} \), where \( C \), \( H \), and \( W \) denote the number of channels, height and width respectively. We first  compute an uncertainty map \( \mathbf{U} \in \mathbb{R}^{H \times W} \) for each spatial position across all channels formulated as
\begin{equation}
\mathbf{U} = \text{Uncertainty}(\mathbf{X})
\end{equation}
where we utilize the standard deviation as our uncertainty metric, a choice validated by the results presented in Section 4.4.2. Specifically, for the input feature map $\mathbf{X}$, we calculate the standard deviation across all channels \( C \) for each spatial position \( (h, w) \) as
\begin{equation}
\mathbf{U}_{h,w} = \sqrt{\frac{1}{C} \sum_{c=1}^C (\mathbf{X}_{c,h,w} - \mu_{h,w})^2}
\end{equation}
where \( \mu_{h,w} \) represents the mean value at that spatial position across all channels.
% \begin{equation}
% \mu_{b,h,w} = \frac{1}{C} \sum_{c=1}^C \mathbf{X}_{b,c,h,w}
% \end{equation}
This calculation captures the pixel-level standard deviation across channels, where higher uncertainty typically corresponds to key regions, such as object boundaries or foreground regions, while lower uncertainty indicates background consistency. By focusing on pixel-level uncertainty, we can more precisely identify key regions for medical image segmentation, which is often critical when identifying pathological regions or organ boundaries.

% \noindent \textbf{Uncertainty-based sorting}.
The uncertainty map \( \mathbf{U} \) is then sorted in descending order, resulting in $\mathbf{U'}$, which ranks the spatial locations from high-uncertainty regions (foreground and boundaries) to low-uncertainty regions (background) defined as
\begin{equation}
\mathbf{U'}, { \textit{Idx}} = \text{Sort}(\mathbf{U})
\end{equation}
This allows the model to prioritize regions with higher complexity or importance during subsequent operations. 

% \noindent \textbf{Feature map rearrangement}.
With the sorted indices \(  \textit{Idx} \), we rearrange the original feature map \( \mathbf{X} \) to create \( \mathbf{X'} \), where regions of high uncertainty are treated intensively, shown as
\begin{equation}
\mathbf{X'} = \text{Rearrange}(\mathbf{X}, \textit{Idx})
\end{equation}
This reorganization prepares the feature map for efficient scanning

\noindent \textbf{Scan Extension Operation}.  
We introduce two primary scanning strategies on the rearranged feature map \( \mathbf{X'} \) to focus on regions with different uncertainty levels, as shown in~\autoref{fig:CD-Block} II: 

1) \textit{Sequential scanning} (\( \text{Scan}^{se} \)): This approach processes spatial locations in ascending or descending order of pixel uncertainty, allowing high-uncertainty areas, such as foreground objects and boundaries, to be modeled with greater density. Sequential scanning ensures that these critical regions are thoroughly captured, enabling the model to focus on areas essential for accurate segmentation. 

2) \textit{Skip scanning} (\( \text{Scan}^{sk} \)): This approach samples spatial locations at regular intervals across the certainty spectrum, promoting interaction between high-uncertainty and low-uncertainty regions. 
% By combining sequential and skip scanning, UD-SSM effectively captures both local and global features.

Combining these two methods, UD-SSM leverages four distinct scanning sequences to comprehensively capture local and global features: sequential and skip scans from high to low uncertainty (\( \mathbf{y_1} \) and \( \mathbf{y_2} \)), and sequential and skip scans from low to high uncertainty (\( \mathbf{y_3} \) and \( \mathbf{y_4} \)), defined as
% Based on $\text{Scan}^{se}$ and $\text{Scan}^{sk}$, UD-SSM utilizes four distinct scanning sequences: sequential and skip scans from high to low uncertainty (\( \mathbf{y_1} \) and \( \mathbf{y_2} \)), and sequential and skip scans from low to high uncertainty (\( \mathbf{y_3} \) and \( \mathbf{y_4} \)), shown as
\begin{align}
\mathbf{y_1} = \text{Scan}^{se}_{high \to low}(\mathbf{X'}) \\
\mathbf{y_2} = \text{Scan}^{sk}_{high \to low}(\mathbf{X'}) \\
\mathbf{y_3} = \text{Scan}^{se}_{low \to high}(\mathbf{X'}) \\
\mathbf{y_4} = \text{Scan}^{sk}_{low \to high}(\mathbf{X'}) 
\end{align}

To enhance the advantages of low-to-high uncertainty scanning and adjust the contribution of each individual scanning sequence, we introduce learnable weights (\( \mathbf{\alpha_1}, \mathbf{\alpha_2}, \mathbf{\alpha_3}, \mathbf{\alpha_4} \)) to dynamically adjust the weight of each scanning path according to its effectiveness in capturing the key area
\begin{equation}
\mathbf{y'_i} = \mathbf{y_i} \cdot \mathbf{\alpha_i} \quad \text{for} \quad i = 1, 2, 3, 4
\end{equation}

\noindent \textbf{S6 block \& Recovery processing}.  
The reweighted features from the four scanning sequences are processed by the S6 block~\cite{gu2023mamba}. Finally, a recovery step consolidates these directional features and restores them to their original spatial configuration, preserving the positional accuracy essential for precise segmentation, defined as
\begin{equation}
\mathbf{y^r_i} = \text{Recover}(\text{S6}(\mathbf{y'_i})) \quad \text{for} \quad i = 1, 2, 3, 4
\end{equation}

The final output of UD-SSM, denoted as \( \mathbf{y}_{\text{UD-SSM}} \), is the sum of these recovered features
\begin{equation}
\mathbf{y}_{\text{UD-SSM}} = \sum_{i=1}^{4} \mathbf{y^r_i}
\end{equation}

\subsection{Objective Function}\label{sec:3.42}

To improve the performance of high-to-low uncertainty scanning in the decoding stage and improve the overall segmentation results, we introduce a cosine consistency constraint at the end of the decoder. This constraint is applied between sequential and skip scans performed in both directions (from high-to-low and low-to-high uncertainty). By aligning the results from low-to-high uncertainty scans with those from high-to-low uncertainty scans, we ensure consistency in feature representation across different scanning directions. The cosine consistency loss is defined as
\begin{equation}
L_{\textit{cos}} = 1 - \frac{{\textit{cos\_sim}}(\mathbf{y_1^r}, \mathbf{y_3^r}) + {\textit{cos\_sim}}(\mathbf{y_2^r}, \mathbf{y_4^r})}{2}
\end{equation}
where \( \textit{cos\_sim} \)$(\cdot)$ represents the average cosine similarity between the forward and backward sequential and skip scans. By maximizing this similarity, we aim to minimize discrepancies between the two scanning directions, thereby reinforcing the consistency of the final segmentation outputs.

Finally, the overall loss function combines the supervised loss $L_{\textit{sup}}$  with the cosine consistency loss $L_{\textit{cos}}$:
\begin{equation}
L = L_{\textit{sup}} + \lambda L_{\textit{cos}}
\end{equation}
where \( L_{\textit{sup}} \) represents the combined cross-entropy and Dice loss and \( \lambda \) is a hyperparameter that balances these two components.

\label{sec:method}

\section{Experiments}

\subsection{Dataset}
We evaluated the  performance of the proposed UD-Mamba on three medical image datasets: 

\noindent \textbf{The DigestPath dataset}~\cite{da2022digestpath} comprises whole slide images (WSIs) for tumor lesions segmentation in colonoscopy. We randomly divided 130 malignant WSIs into three subsets: 100 for training, 10 for validation, and 20 for testing. For model training, WSIs were further partitioned into 256 × 256 pixel patches, yielding a training set of 29,412 patches. Our model evaluation was conducted at the WSI level.

\noindent \textbf{The ISIC 2018 dataset}~\cite{codella2019skin} is a public dataset for skin lesion segmentation containing 2,694 dermoscopy images with corresponding pixel-level labels. We follow VM-UNet~\cite{ruan2024vm} to split the dataset into training and test sets.

\begin{table*}[tbp]
  \centering
  \caption{Performance comparison of different networks on DigestPath and ISIC 2018 datasets.}
  \scalebox{0.7}{
    \begin{tabular}{c|ccccc|ccccc}
    \toprule
    \multicolumn{1}{c|}{Dataset} & \multicolumn{5}{c|}{\textbf{ISIC 2018}} & \multicolumn{5}{c}{\textbf{DigestPath}} \\
    \midrule
    Network & DSC(\%)\(\uparrow\) & IoU(\%)\(\uparrow\) & ACC(\%)\(\uparrow\) & Spe(\%)\(\uparrow\) & Sen(\%)\(\uparrow\) & DSC(\%)\(\uparrow\) & IoU(\%)\(\uparrow\) & ACC(\%)\(\uparrow\) & Spe(\%)\(\uparrow\) & Sen(\%)\(\uparrow\) \\
    \midrule
    UNet~\cite{ronneberger2015u} & 86.51 & 77.81 & 92.91 & 94.90 & 88.69 & 77.96 & 64.91 & 94.30 & 96.09 & 80.74 \\
    UNet++~\cite{zhou2019unet++} & 87.36 & 79.20 & 93.10 & 95.59 & 88.71 & 78.37 & 65.43 & 94.52 & 96.13 & 80.41 \\
    Att-UNet~\cite{oktay2018attention} & 87.47 & 79.31 & 93.12 & 95.77 & 88.83 & 78.28 & 65.24 & 94.38 & 95.78 & 81.57 \\
    TransUNet~\cite{chen2024transunet} & 88.12 & 80.32 & 93.91 & 94.04 & 89.40 & 79.30 & 66.74 & 94.64 & 96.27 & 81.18 \\
    SwinUNet~\cite{cao2022swin} & 87.20 & 79.27 & 93.49 & 96.22 & 87.30 & 79.15 & 66.54 & 94.75 & 96.84 & 79.98 \\
    H2Former~\cite{he2023h2former} & 87.62 & 79.56 & 94.08 & 95.13 & 88.52 & 79.61 & 67.23 & 94.47 & 95.40 & 81.35 \\
    Mamba-UNet~\cite{wang2024mamba_unet} & 87.86 & 80.36 & 93.79 & 96.36 & 89.61 & 79.92 & 67.41 & 94.65 & 96.06 & 82.47 \\
    Swin-Umamba~\cite{liu2024swin} & 88.60 & 81.12 & 94.53 & 95.87 & 89.21 & 79.45 & 67.32 & 94.60 & 96.10 & 82.55 \\
    \rowcolor{lightgray}
    Ours & 89.15 & 81.94 & 94.60 & 96.26 & 89.55 & 80.89 & 68.64 & 94.98 & 96.44 & 83.34 \\
    \bottomrule
    \end{tabular}%
    }
  \label{tab:ISIC_tissue}%
\end{table*}

\noindent \textbf{The ACDC dataset}~\cite{bernard2018deep} consists of cardiac cine MRI scans from 100 patients, used for the segmentation of three cardiac substructures: the Left Ventricle (LV), Right Ventricle (RV), and Myocardium (MYO). We split the dataset into 70\% for training, 10\% for validation, and 20\% for testing. All slices were resized to a uniform resolution of 256 × 256 pixels before training. 

For both DigestPath and ISIC 2018 datasets, we conducted a detailed evaluation using performance metrics including mean Intersection over Union (mIoU), Dice Similarity Coefficient (DSC), Accuracy (Acc), Sensitivity (Sen), and Specificity (Spe). For the ACDC dataset, performance was evaluated using the DSC, mIoU, and 95\% Hausdorff Distance (HD$_{95}$). Given the fixed anatomical structures in ACDC, the inclusion of HD$_{95}$ provides a more robust assessment of boundary accuracy.

\subsection{Implementation details}\label{sec:4.2}
All experiments were conducted using the PyTorch framework on an Ubuntu desktop equipped with an NVIDIA RTX A6000 GPU. The training was performed using Stochastic Gradient Descent (SGD) with a multi-step learning rate strategy, initially set to 0.01. The total number of training epochs was fixed at 300. For UD-Mamba, each layer of both the encoder and decoder corresponds to two UD blocks. We utilize weights pre-trained on ImageNet-1K~\cite{deng2009imagenet} to initialize the encoder. The value of $\lambda$ is 0.3. More implementation details are in the supplementary materials.

\subsection{Comparison with advanced methods}
We compared our UD-Mamba with state-of-the-art medical image segmentation methods including CNN-based approaches (UNet~\cite{ronneberger2015u}, UNet++~\cite{zhou2019unet++} and Att-UNet~\cite{oktay2018attention}), Transformer-based models (TransUNet~\cite{chen2024transunet}, SwinUNet~\cite{cao2022swin} and H2Former~\cite{he2023h2former}), as well as Mamba-based models (Mamba-UNet~\cite{wang2024mamba_unet} and Swin-Umamba~\cite{liu2024swin}).

\begin{table}[tbp]
  \centering
    \caption{Comparison of different networks on ACDC dataset. Here, `RV', `MYO' and `LV' mean the DSC for segmenting different cardiac substructures.}
    \scalebox{0.9}{
        \begin{tabular}{c|c|ccc|cc}
        \toprule
        Network & DSC (\%) $\uparrow$ & RV & MYO & LV & mIoU (\%) $\uparrow$ & $HD_{95}$ (mm) $\downarrow$ \\
        \midrule
       UNet~\cite{ronneberger2015u}  & 90.07 & 89.11 & 87.22 & 93.89 & 82.42 & 2.74 \\
        UNet++~\cite{zhou2019unet++} & 90.23 & 89.08 & 87.65 & 93.96 & 82.64 & 1.90 \\
        Att-UNet~\cite{oktay2018attention} & 89.17 & 88.45 & 86.14 & 92.94 & 81.04 & 3.17 \\
        TransUNet~\cite{chen2024transunet} & 90.70 & 91.71 & 87.74 & 92.68 & 83.50 & 2.76 \\
        SwinUNet~\cite{cao2022swin} & 89.45 & 90.52 & 86.23 & 91.60 & 81.55 & 3.56 \\
        H2Former~\cite{he2023h2former} & 90.93 & 90.30 & 88.21 & 94.28 & 83.78 & 2.74 \\
        Mamba-UNet~\cite{wang2024mamba_unet} & 91.08 & 90.80 & 88.09 & 94.35 & 84.03 & 1.40 \\
        Swin-Umamba~\cite{liu2024swin} & 89.80 & 88.37 & 87.65 & 93.39 & 82.09 & 2.98 \\
        \rowcolor{lightgray}
        Ours  & 91.99 & 90.85 & 90.69 & 94.45 & 85.48 & 1.31 \\
        \bottomrule
        \end{tabular}
    }
    \label{tab:ACDC}
\end{table}

\autoref{tab:ISIC_tissue} shows quantitative results on the ISIC 2018 and DigestPath datasets. Our UD-Mamba significantly outperforms CNN-based approaches. Specifically, UD-Mamba achieved improvements of 1.68\% and 2.52\% in DSC over the best CNN methods on the ISIC 2018 and DigestPath datasets, respectively. Moreover, the mIoU scores respectively increased by 2.63\% and 3.21\%. Compared to Transformer-based models such as TransUNet~\cite{chen2024transunet}, our method demonstrated a notable advantage in mIoU, with increases of 1.62\% for ISIC 2018 and 1.90\% for DigestPath. Additionally, when compared to the representative Mamba-based model Mamba-UNet~\cite{wang2024mamba_unet}, UD-Mamba improved the mIoU by 1.58\% and 1.23\% on the two datasets, respectively.

\begin{table}[tbp]
    \centering
    \caption{Comparison of different methods for measuring model complexity and efficiency on the ACDC dataset.}
    \scalebox{0.9}{
        \begin{tabular}{c|c|c|c}
        \toprule
        \textbf{Category} & \textbf{Method} & \textbf{Parameters} $\downarrow$ & \textbf{FLOPs} $\downarrow$ \\
        \toprule
        \multirow{1}{*}{CNN-based} & UNet++~\cite{ronneberger2015u} & \textbf{9.16M} & 34.75G \\
        \midrule
        \multirow{3}{*}{Transformer-based} & TransUNet~\cite{chen2024transunet} & 105.32M & 38.57G \\
        & SwinUNet~\cite{cao2022swin} & 40.32M & 25.63G \\
        & H2Former~\cite{he2023h2former} & 33.68M & 32.46G \\
        \midrule
        \multirow{2}{*}{Mamba-based} & SwinUmamba~\cite{liu2024swin} & 59.88M & 40.94G \\
        & Ours & 19.12M & \textbf{5.91G} \\
        \bottomrule
        \end{tabular}
    }
    \label{tab:flops}
\end{table}

\begin{figure}
    \centering
    \includegraphics[width=1\linewidth]{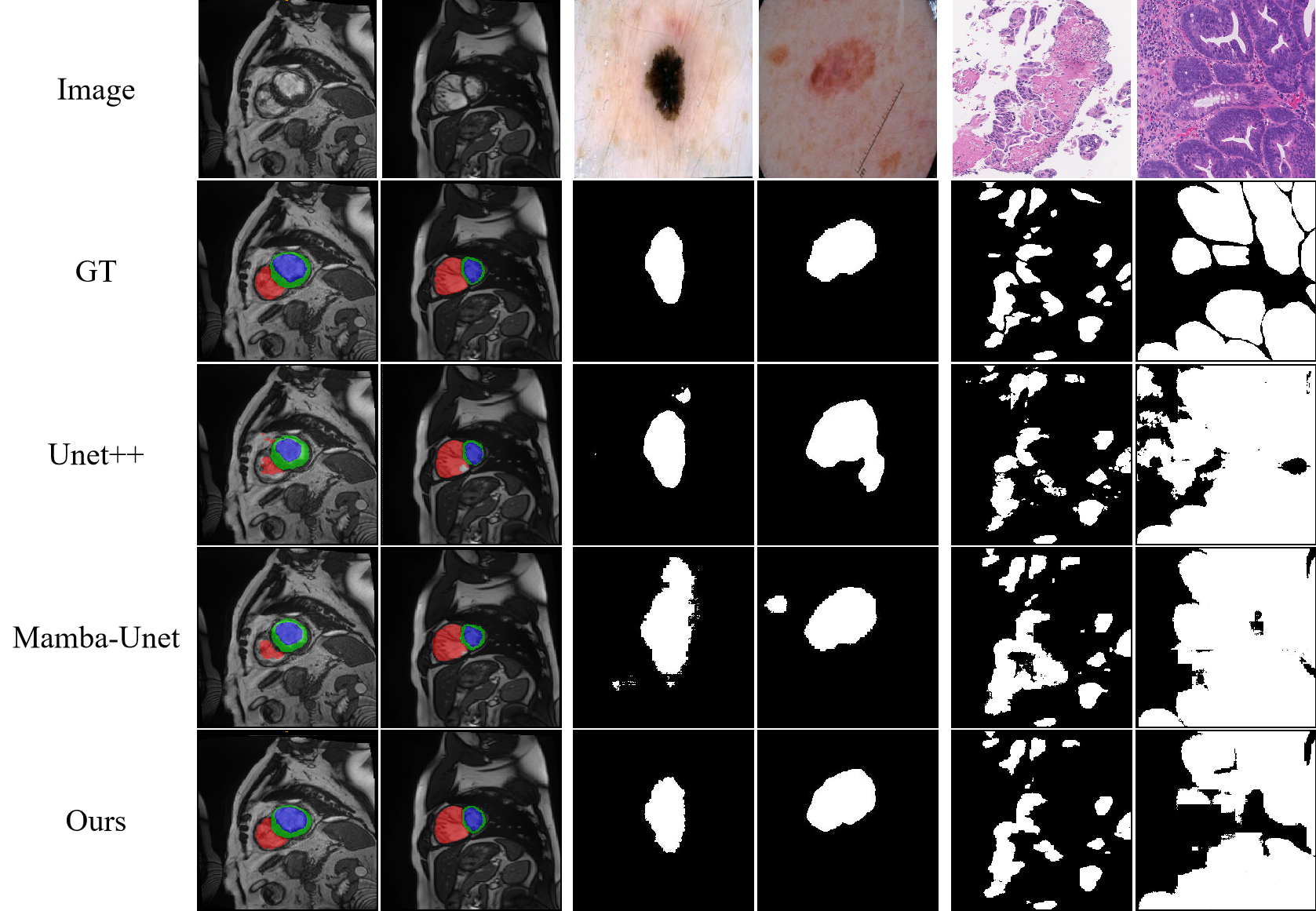}
    \caption{Visual comparisons of segmentation results from UD-Mamba and various other methods are conducted across three different datasets.}
    \label{fig:Visual}
\end{figure}

For the ACDC dataset, \autoref{tab:ACDC} presents a comparison of results with other methods. Compared to the best-performing Mamba-UNet~\cite{wang2024mamba_unet}, our approach demonstrated significant improvements, with increases of 0.91\% and 1.45\% in DSC and mIoU, respectively, while reducing the HD$_{95}$ metric to 1.31 mm. 
The results in \autoref{tab:flops} also demonstrate that our model significantly reduces the parameter count compared to Transformer-based and Mamba-based networks and achieves the lowest FLOPs among all networks. This highlights its ability to deliver strong segmentation performance while optimizing computational efficiency.
The visualization results on three datasets are shown in \autoref{fig:Visual}.

%\begin{figure}
    %\centering
    %\includegraphics[width=0.8\linewidth]{vis3.pdf}
    %\caption{Visual comparisons of uncertainty maps.}
    %\label{fig:compare_pic}
%\end{figure}

\subsection{Ablation studies}
We conducted experimental verification on the DigestPath dataset~\cite{da2022digestpath} in the ablation study section.

\subsubsection{Uncertainty scanning strategies}
%We conducted ablation experiments on the pixel-level channel uncertainty-driven scanning operation and its optimization strategies. As shown in~\autoref{tab:Uncertainty}, without our components (first row), the method degenerates into a standard position-based scanning approach. We observed that the uncertainty-driven scanning method yielded superior results compared to the original scanning method (DSC: 80.32\% vs. 79.92\%). This confirms that the uncertainty-driven scanning approach effectively separates uncertain regions representing foreground and boundaries from background-related areas, thereby enhancing local modeling capabilities. %As illustrated in~\autoref{fig:compare_pic}, our method demonstrates excellent modeling capability for target regions compared to the traditional position-based scanning method used in Mamba-UNet~\cite{wang2024mamba_unet}. 
%Furthermore, the re-weighting and consistency constraint strategies further enhance the model's representational capacity. These strategies amplify the advantages of scanning from high to low uncertainty while mitigating the limitations of scanning from low to high uncertainty, resulting in an improved DSC of 80.89\%.

We evaluated the impact of individual pixel-level uncertainty-driven scanning components (\( y_1, y_2, y_3, y_4 \)) and optimization strategies (re-weighting and \( L_{\text{cos}} \)) in the UD-SSM framework. \autoref{tab:Uncertainty} summarizes the segmentation performance across different configurations, highlighting the effectiveness of each component and strategy.

The results indicated that combining scanning strategies yielded better performance than using individual components alone. For instance, using the strategy \( y_4 \) resulted in a DSC of 76.87\%. However, combining forward and backward bidirectional scanning strategies, specifically the skip scanning components (\( y_2+y_4 \)), improved the DSC to 77.90\%. Similarly, sequential scanning produced consistent results (\( y_1+y_3 \): 76.27\% DSC vs. \( y_3 \): 75.67\% DSC). The integration of all four components (\( y_1+y_2+y_3+y_4 \)) achieved a higher DSC of 80.32\%, surpassing both sequential and skip scanning alone and outperforming the position-based sequential scanning in the vanilla Mamba model (the first row). This demonstrates the complementary advantages of integrating both sequential and skip scanning to capture both local and global features.

Furthermore, the addition of optimization strategies further improved performance. Re-weighting alone increased the DSC from 80.32\% to 80.72\%, while incorporating the cosine similarity constraint (\( L_{\text{cos}} \)) boosted the DSC to 80.89\%, with the mIoU of 68.64\% and accuracy of 94.98\%. This combination optimized the contribution of the scanning strategies and aligned features across directions, resulting in superior segmentation accuracy.

\begin{table}[tbp]
    \centering
    \caption{Comparison of the effects of uncertainty scanning components ($y_1$, $y_2$, $y_3$, $y_4$) and optimization strategies.}
    \scalebox{0.8}
    {
        \begin{tabular}{cccc|cc|ccc}
            \toprule
            \multicolumn{4}{c|}{\textbf{UD-SSM Components}} & \multicolumn{2}{c|}{\textbf{Optimization Strategies}} & \multicolumn{3}{c}{\textbf{Performance Metrics}} \\
            \hline
            $y_1$ & $y_2$ & $y_3$ & $y_4$ & Reweight & $L_{\text{cos}}$ & \textbf{DSC} $\uparrow$ & \textbf{mIoU} $\uparrow$ & \textbf{ACC} $\uparrow$ \\
            \hline
             -- & --  & --  & --  & --  & --  & 78.15 & 65.93 & 93.95 \\
             &  & \checkmark &  &  &  & 75.67 & 62.98 & 94.42 \\
             &  &  & \checkmark &  &  & 76.87 & 64.35 & 64.67 \\
            \checkmark &  & \checkmark &  &  &  & 76.27 & 63.40 & 94.10 \\
             & \checkmark &  & \checkmark &  &  & 77.90 & 64.90 & 94.37 \\
            \checkmark & \checkmark & \checkmark & \checkmark &  &  & 80.32 & 68.14 & 94.93 \\
            \hline
            \checkmark & \checkmark & \checkmark & \checkmark & \checkmark &  & 80.72 & 68.55 & 94.97 \\
            \checkmark & \checkmark & \checkmark & \checkmark & \checkmark & \checkmark & $\textbf{80.89}$ & $\textbf{68.64}$ & $\textbf{94.98}$ \\
            \bottomrule
        \end{tabular}
    }
    \label{tab:Uncertainty}
\end{table}

In conclusion, \autoref{tab:Uncertainty} demonstrates that the complete integration of scanning components with re-weighting and consistency constraints maximized the model's ability to handle uncertain regions, validating the effectiveness of UD-SSM for uncertainty-driven segmentation.

% need to discuss: 
% y1+y3 > y3 or y2+y4 > y4
% y1+y3+y2+y4 ? use each alone
% Reweight is useful
% consistency constraint is useful

%\begin{table}[tbp]
    %\centering
    %\caption{Comparison of the effects of uncertainty scanning and its optimization strategies.}
    %\scalebox{0.85}{
        %\begin{tabular}{ccc|ccc}
            %\toprule
            %\textbf{UD-SSB} & \textbf{Reweight} & \textbf{$L_{\text{cos}}$} & \textbf{DSC} $\uparrow$ & \textbf{mIoU} $\uparrow$ & \textbf{ACC} $\uparrow$ \\
            %\hline
             %&  &  & 79.92 & 67.41 & 94.65 \\
            %\checkmark &  &  & 80.32 & 68.14 & 94.93 \\
            %\checkmark &  & \checkmark & 80.41 & 68.18 & 94.94 \\
            %\checkmark & \checkmark &  & 80.72 & 68.55 & 94.97 \\
            %\checkmark & \checkmark & \checkmark & $\textbf{80.89}$ & $\textbf{68.64}$ & $\textbf{94.98}$ \\
            %\bottomrule
        %\end{tabular}
    %}
    %\label{tab:Uncertainty}
%\end{table}

\begin{table}[tbp]
    \centering
    \caption{Comparison of different methods for measuring the uncertainty of channels.}
    \scalebox{0.85}{
        \begin{tabular}{c|ccc}  
        \toprule
           \textbf{Method} & \textbf{DSC} $\uparrow$ & \textbf{mIoU} $\uparrow$ & \textbf{ACC} $\uparrow$ \\
            \toprule
            Mad &  80.75 & 68.58 & 94.97  \\
            Range & 79.89 & 68.07 & 94.88  \\
            Entroph & 80.28 & 67.96 & 94.89  \\
            Variance & 80.02 & 67.26 & 94.75  \\
            STD & $\textbf{80.89}$ & $\textbf{68.64}$ & $\textbf{94.98}$\\
            \bottomrule
        \end{tabular}
    }
    \label{tab:uncertain_method}
\end{table}

\subsubsection{Different uncertainty calculation methods}
To evaluate various criteria for measuring channel uncertainty, we conducted ablation experiments. These criteria include Mean Absolute Deviation (MAD), Standard Deviation (STD), Variance, Entropy and the Range between the two highest values. As illustrated in~\autoref{tab:uncertain_method}, the use of STD provides a stable measure of data dispersion. This stability enables the model to more reliably identify true regions of uncertainty, rather than being misled by noise or outliers. Consequently, the method that employs STD  to calculate uncertainty achieved the best results, attaining the highest DSC of 80.89\%, the highest mIoU of 68.64\% and the highest ACC of 94.98\%.

\subsubsection{Uncertainty calculation region}
To evaluate the effectiveness of pixel-level uncertainty-driven scanning in scenarios lacking explicit spatial features, we conducted comparative experiments focusing on the size of the regions used for uncertainty calculations. Instead of relying solely on the uncertainty of individual pixels, we extended the calculation to larger regions to retain some degree of spatial information. These regions are defined as uncertainty blocks with dimensions \(a \times a\). 

\begin{table}
    \centering
  \caption{Comparison of different methods for calculating uncertainty region.}
   \scalebox{1}{
    \begin{tabular}{c|cccc|cc}
    \toprule \multirow{1.8}{*}{Size} & \multicolumn{4}{c|}{ Static } & \multicolumn{2}{c}{ Dynamic } \\
   \cline { 2 - 7 } & 1   & 2   & 4   & 8 & $a_{v}/a_{v}^{min}$  & $a_{v}^{max}/a_{v}$  \\
    \hline
    DSC \(\uparrow\)  & 80.89  & 80.44  & 80.19  & 79.82 & 79.85  & 79.93 \\
    \bottomrule
    \end{tabular}}%
  \label{tab:size}%
\end{table}

Our experimental design explores both fixed and dynamically adjusted values for \(a\). For fixed-size regions, we varied \(a\) from 1 up to \(a_{v}^{min}\). In the case of dynamically adjusted regions, two strategies were employed: (1) proportional scaling, where \(a = a_{v} / a_{v}^{min}\), allowing \(a\) to increase proportionally with the feature vector size \(a_v\); and (2) inverse proportional scaling, where \(a = a_{v}^{max} / a_v\), causing \(a\) to decrease as the feature vector size \(a_v\) increases. Here, \(a_v\) refers to the feature vector size at each stage before entering the UD-SSM, \(a_{v}^{max}\) represents the feature vector size upon the first entry into the UD-SSM, and \(a_{v}^{min}\) denotes the feature vector size at the bottleneck layer. In UD-Mamba, \(a_{v}^{max}\) and \(a_{v}^{min}\) are set to 64 and 8, respectively. After calculating the average uncertainty value for each region, these values are used to rank the regions for subsequent scanning. As demonstrated in~\autoref{tab:size}, pixel-level uncertainty-driven scanning consistently outperforms both dynamic and static region-based methods. This result highlights the advantages of pixel-level granularity in determining uncertainty for fine-grained tasks like medical image segmentation. Compared to broader region-based uncertainty approaches, pixel-level uncertainty focuses on capturing local variations, providing a more precise method for identifying critical segmentation targets.

\subsubsection{Analysis for re-weighting values}
\begin{figure}
    \centering
    \includegraphics[width=1\linewidth]{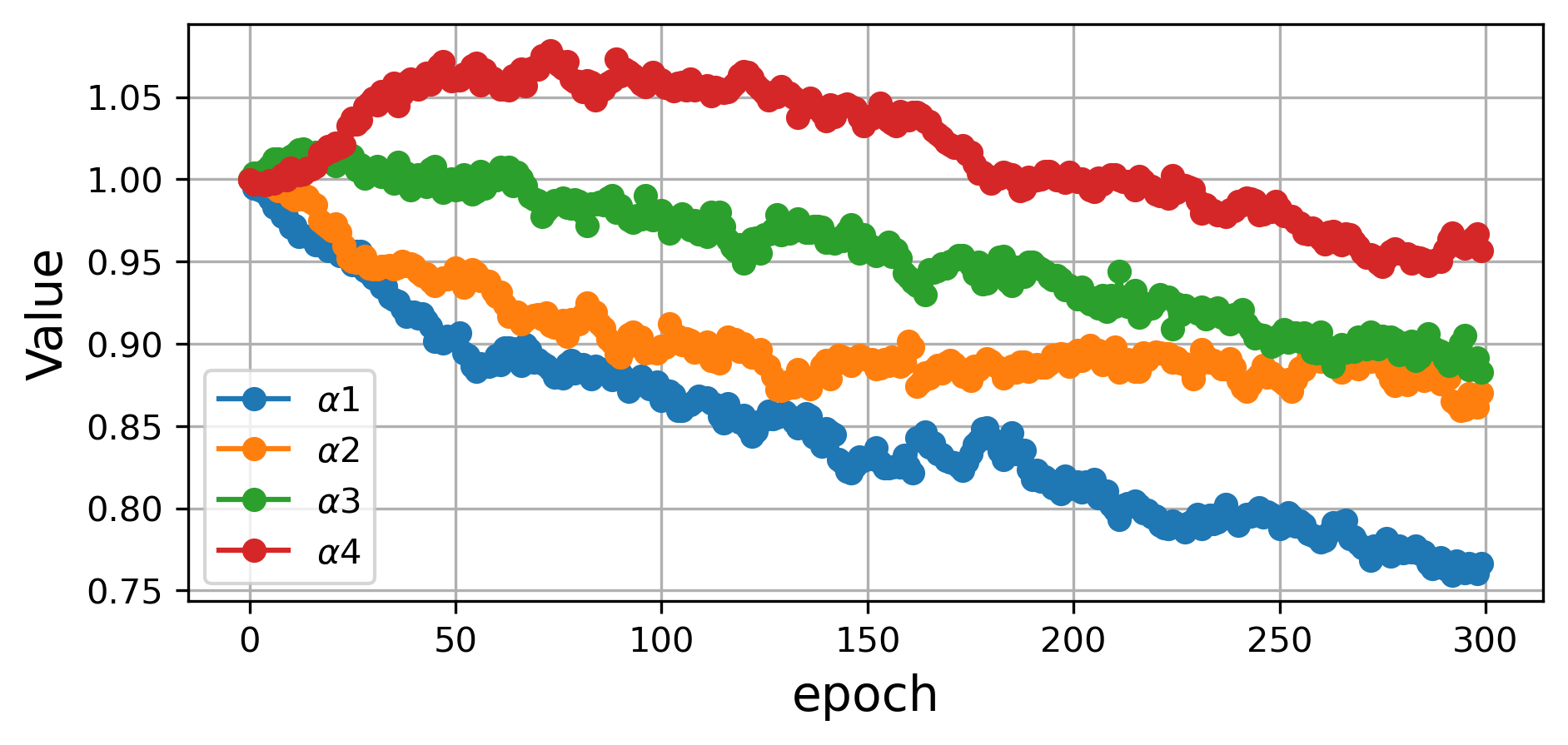}
    \caption{Analysis of recorded values for four learnable reweighting parameters.}
    \label{fig:epock}
\end{figure}

\autoref{fig:epock} illustrates the evolution of the four learnable parameters $\alpha_1$, $\alpha_2$, $\alpha_3$, and $\alpha_4$, which reweight the four different scanning sequences, throughout the training process. All four parameters exhibit a downward trend, with $\alpha_3$ and $\alpha_4$ showing a less pronounced decrease compared to $\alpha_1$ and $\alpha_2$. This pattern suggests that during training, the scanning processes from high to low uncertainty levels, corresponding to $\alpha_3$ and $\alpha_4$, contribute more significantly than the scanning processes from low to high uncertainty levels associated with $\alpha_1$ and $\alpha_2$. This observation indirectly corroborates the conclusion proposed in~\autoref{fig:compare}.

%\subsubsection{Ablation of hyperparameter $\lambda$}
%For the hyperparameter $\lambda$, which controls the magnitude of the consistency constraint loss between bidirectional scans, we conducted ablation experiments to determine its optimal range. As shown in~\autoref{fig:lamda}, the best results were obtained when $\lambda$ was set to 0.3.

%\begin{figure}
    %\centering
    %\includegraphics[width=1\linewidth]{lamda3.png}
    %\caption{Sensitivity analysis of the hyperparameter $\lambda$.}
    %\label{fig:lamda}
%\end{figure}

\label{sec:Experiments}

\section{Conclusion}
In this paper, we introduce UD-Mamba, a novel architecture designed to address Mamba's limitations in local feature modeling. By integrating a pixel-level channel uncertainty-driven mechanism, UD-Mamba effectively prioritizes pixels based on channel uncertainty, enabling comprehensive and efficient feature extraction. Furthermore, as scanning from low-uncertainty to high-uncertainty vectors typically yields greater benefits than the reverse process, we introduce four learnable parameters to explore the impact of various scanning sequences on the autoregressive Mamba framework. Concurrently, we enhance the efficacy of transitions from high-uncertainty to low-uncertainty regions by constraining the cosine similarity loss between forward and backward scanning results. Experimental results on three medical imaging datasets demonstrate UD-Mamba's superior performance in medical image segmentation tasks compared to traditional models.
Future work will focus on developing more precise and effective uncertainty estimation methods, as model performance depends heavily on accurate channel uncertainty estimation.  Additionally, we aim to expand the application of UD-Mamba to a wider range of medical image segmentation challenges.

\bibliographystyle{splncs04}
\bibliography{mybibliography}

\end{document}